\documentclass[aps,pra,showpacs,twocolumn]{revtex4}
\usepackage{graphicx}
\usepackage{bm}
\usepackage{amsfonts}

\usepackage{amssymb,amsthm,dsfont,bm,color,ulem}

\begin{document}

\title{Phase-number uncertainty from Weyl commutation relations}

\author{Alfredo Luis and Gonzalo Donoso}
\affiliation{Departamento de \'{O}ptica, Facultad de Ciencias
F\'{\i}sicas, Universidad Complutense, 28040 Madrid, Spain }

\begin{abstract}
We derive suitable uncertainty relations for characteristics functions of phase and number 
variables obtained from the Weyl form of commutation relations. This is applied to 
finite-dimensional spin-like systems, which is the case when describing the phase difference 
between two field modes, as well as to the phase and number of a single-mode field.
Some contradictions between the product and sums of characteristic functions are noted. 
\end{abstract}


\maketitle

\section{Introduction}

Uncertainty relations are a key item of the quantum theory. This is from fundamental reasons, but also
regarding practical applications, since phase-number uncertainty relations are the heart of the quantum 
limits to the precision of signal detection schemes \cite{HL}.

Typically, uncertainty relations are expressed in terms of variances and are derived directly  from the 
Heisenberg form of commutation relations. However, this approach is not always useful. On the 
one hand, variance may not be a suitable uncertainty measure. This is specially clear regarding 
periodic phase-angle variables \cite{OP}.  On the other hand, the phase may not admit a simple 
well-behaved operator description suitable to obey a Heisenberg form of commutation relations 
with the number operator \cite{GWG,AE,SG}. This has lead to the introduction of alternative 
uncertainty relations \cite{AE,SG,H,LP,OP2,MP}, some of them involving characteristic functions \cite{LU03}.

In this regard, a recent work has proposed an uncertainty relation for position and momentum  
based on characteristic functions, which is derived directly from the Weyl form of commutation relations 
\cite{RTW}. In this work we translate this approach to phase-number variables. Despite the problems 
that quantum phase encounters, a very fundamental approach admits without difficulties the Weyl 
form of commutation relations and has well-defined characteristic functions. Therefore, the approach 
in Ref. \cite{RTW} is a quite interesting formulation particularly suited to phase-angle variables. We 
also show that this encounters fundamental ambiguities when contrasting different slightly different 
alternative implementations, as it also holds for other approaches \cite{BPP,PL,LBP}.

Let us point out that the Weyl form is equivalent to say that every system state experiences a 
global phase shift after a cyclic transformation in the corresponding phase space. This implies 
that the quantum structure including uncertainty relations might be traced back to a geometric 
phase \cite{LU01}. 

\section{Spin-like systems}

Let us consider general systems describable in a finite-dimensional space as a spin $j$. This 
admits very general scenarios, including especially the phase difference between two modes 
of the electromagnetic field. This is because the total number of photons $N$ is compatible with 
the phase difference and defines finite-dimensional subspaces of dimension $N+1$, where 
$N$ plays the role of the spin modulus as  $j=N/2$ \cite{PD}. 
 
Let us focus on a spin component $j_3$ and the canonically conjugate phase  $\phi$. To avoid 
periodicity problems we focus on the complex exponential of $\phi$, we shall call $E$, this is 
$E= e^{i \phi}$. The eigenvectors $E | \tilde{m} \rangle = e^{i \frac{2 \pi}{2j+1} \tilde{m} } | \tilde{m} \rangle$ 
can be referred to as  phase states \cite{PDT}, being
\begin{equation}
| \tilde{m} \rangle = \frac{1}{\sqrt{2j+1}} \sum_{m=-j}^j e^{ -i \frac{2 \pi}{2j+1} m \tilde{m} }
| m \rangle , 
\end{equation}
where $| m \rangle$ are the eigenstates of $j_3$, as usual $j_3 | m \rangle = m | m \rangle$, and 
$m,\tilde{m} = -j,-j+1,\ldots , j$. Likewise, we may define the exponential of $j_3$ as 
\begin{equation}
F= e^{ i \frac{2 \pi}{2j+1}  j_3} . 
\end{equation}
These exponentials $E$ and $F$  are quite suited to the Weyl form of commutation 
relation \cite{WF}
 \begin{equation}
 \label{Ws}
 E^k F^\ell =   e^{-i \frac{2 \pi}{2j+1} k \ell} F^\ell E^k  ,
 \end{equation}
for any $k,\ell \in \mathbb{Z}$. It is worth noting that the Weyl form has a quite interesting meaning 
when expressed as 
 \begin{equation}
E^{\dagger k} F^{\dagger \ell} E^k F^\ell =   e^{ -i \frac{2 \pi}{2j+1} k \ell}  .
 \end{equation}
This represents a cyclic transformation in the form of a closed excursion over a $k \times \ell$ rectangle 
in the associated phase space for the problem.  The result is that every system state acquires a 
global phase after returning to the starting point. 

Following Ref. \cite{RTW} we can construct the Gram matrix $G$ for the following three vectors 
\begin{equation}
|\Psi \rangle, \qquad  F^\ell |\Psi \rangle, \qquad E^k |\Psi \rangle , 
\end{equation}
where $|\Psi \rangle$ is an arbitrary state assumed pure for simplicity and without loss of generality, 
so that 
\begin{equation}
G = \pmatrix{1 & \Phi & \tilde{\Phi} \cr \Phi^\ast & 1 & \Omega \cr \tilde{\Phi}^\ast & \Omega^\ast & 1} ,
\end{equation}
involving the characteristic functions 
\begin{equation}
\Phi = \langle  \Psi | F^\ell | \Psi \rangle, \quad \tilde{\Phi} = \langle \Psi | E^k | \Psi \rangle, 
\end{equation}
and 
\begin{equation}
\Omega = \langle \Psi | F^{\dagger \ell} E^k | \Psi \rangle,
\end{equation}
which is the term invoking the Weyl commutator (\ref{Ws}).
After the positive semi-definiteness of $G$ we get
\begin{equation}
\det G = 1 - \left | \Phi \right |^2 - \left | \tilde{\Phi} \right |^2 - \left | \Omega \right |^2 + \Theta + \Theta^\ast
\geq 0 ,
\end{equation}
where $\Theta = \Omega \Phi \tilde{\Phi}^\ast$. From this point we can follow exactly the same steps in 
Ref. \cite{RTW} . These involve to construct another Gram matrix after replacing $k$ by $-k$ and $\ell$ by 
$-\ell$, adding the two determinants, using Eq. (\ref{Ws}) and then following some clever simple algebraic 
bounds. This leads to
\begin{equation}
\label{URs}
\left | \Phi \right |^2 + \left | \tilde{\Phi} \right |^2 \leq {\cal B} ,
\end{equation}
with
\begin{equation}
\label{B}
{\cal B} = 2 \sqrt{2} \frac{\sqrt{2}-\sqrt{1-\cos \gamma}}{1+\cos \gamma}, \quad \gamma = \frac{2\pi}{2j+1} k \ell .
\end{equation}
Therefore, most of the analysis and results found in Ref. \cite{RTW} could be translated here. Even, the limit 
of vanishing argument of the characteristic functions may be reproduced in the limit of very large $j$. 

Besides the sums, uncertainty relations can be also formulated as the products of uncertainty estimators. 
In our case from Eq. (\ref{URs}) we can readily derive a bound for the product of characteristic functions 
\begin{equation}
\label{URp}
\left | \Phi \right | \left | \tilde{\Phi} \right | \leq {\cal B}/2 .
\end{equation}
A rather interesting point is that this can lead to conclusions fully opposite to the sum relation (\ref{URs}). 
This is specially so regarding the minimum uncertainty states, as we shall clearly show by some examples below.

The smallest value for the bound ${\cal B}$ is obtained for $\gamma = \pi$. In such a case, the sum of the two 
Gram matrices commented above leads directly to the uncertainty relation 
\begin{equation}
\label{URt}
\left | \Phi \right |^2 + \left | \tilde{\Phi} \right |^2 + \left | \Omega \right |^2 \leq 1,
\end{equation}
where the $\Omega$ term is expressing phase-number correlations that in standard variance-based approaches 
is expressed by the the anti-commutator. If this correlation term is ignored we get the more plain relations:
\begin{equation}
\left | \Phi \right |^2 + \left | \tilde{\Phi} \right |^2 \leq 1, \quad
\left | \Phi \right | \left | \tilde{\Phi} \right | \leq 1/2.
\end{equation}

Let us note that these relations might be called \textit{certainty} instead of \textit{uncertainty} relations since 
we get upper bounds for characteristic functions, that take their maximum value when there is full certainty 
about the corresponding variable.

 \subsection{Example: qubit}
 
The most simple and illustrative example is provided by the case $j=1/2$. The most general state is of the 
form $\rho =  ( \sigma_0 +  \bm{s} \cdot \bm{\sigma} )/2$ where $\bm{\sigma}$ are the Pauli matrices, 
$\sigma_0$ is the identity, and $\bm{s} = \mathrm{tr} ( \rho \bm{\sigma}) $ is a three-dimensional real vector 
with $|\bm{s} | \leq 1$. We can chose the basis so that $F= \sigma_z$ and $E= \sigma_x$. The only nontrivial 
uncertainty relation holds  for $k=\ell=1$ so that $\gamma = \pi$, ${\cal B} =1$ 
\begin{equation}
\Phi = s_z, \quad \tilde{\Phi} = s_x, \quad \Omega = i s_x s_y s_z ,
\end{equation}
and Eqs. (\ref{URs}), (\ref{URt}), and (\ref{URp}) become, respectively
\begin{equation}
s_x^2 + s_z^2 \leq 1, \quad s_x^2 + s_z^2 + s_x^2 s_y^2 s_z^2 \leq 1, \quad  | s_x s_z| \leq 1/2 .
\end{equation}
Actually, $s_x^2 + s_z^2 \leq 1$ is a well-known duality relation expressing complementarity 
\cite{JSV}. The minimum uncertainty both  for Eqs. (\ref{URs}) and (\ref{URt}) holds for every pure 
state  $|\bm{s} | = 1$ with $s_y=0$. 

Turning our attention to the alternative product of characteristic functions in Eq. (\ref{URp}) we get that 
the minimum uncertainty states are those pure states with $s_y=0$ and  $|s_x| = | s_z | = 1/\sqrt{2}$. On 
the other hand, the states with  $s_y=0$ and  $|s_x| = 0$ or $| s_z | = 0$ are of maximum uncertainty, 
contrary to the predictions of the sum relations (\ref{URs})  and (\ref{URt}).
 
\section{Single-mode phase and number}

Next we address the case of the number and phase for a single field mode. There is always the possibility of 
addressing this from the number and phase difference taking a suitable reference state in one of the modes 
\cite{PD,Ban}, but the direct approach has also its advantages. One of them is that it faces the fact that, roughly 
speaking, there is no phase operator. The exponential of the phase is not unitary, but represented 
instead by the one-sided unitary Susskind-Glogower operator \cite{AE,SG}
\begin{equation}
E = \sum_{n=0}^\infty | n \rangle \langle n+1 | ,\quad E |\phi \rangle = e^{i\phi} | \phi \rangle ,
\end{equation}
where $| n \rangle$ are the eigenstates of the number operator $\hat{n}$, and $|\phi \rangle$  are the phase 
states
\begin{equation}
\label{ps}
|\phi \rangle = \frac{1}{\sqrt{2 \pi}} \sum_{n=0}^\infty e^{i n \phi} | n \rangle .
\end{equation}
The lack of unitary is conveniently expressed as 
\begin{equation}
\label{lu}
E^k E^{\dagger k} = I ,\qquad E^{\dagger k} E^k = I - \hat{\Pi}_k ,
\end{equation}
where $ \hat{\Pi}_k$ is the orthogonal projector on the subspace with less than $k$ photons
\begin{equation}
 \hat{\Pi}_k = \sum_{n=0}^{k-1} | n \rangle \langle n | ,
 \end{equation}
and $I$ is the identity. 

This does not prevent the existence of a proper probability distribution for the phase in any field state
$\rho$. This can be defined thanks to the phase states (\ref{ps}) as $P(\phi) = \langle \phi | \rho | \phi 
\rangle$, that lead to the characteristic function
\begin{equation}
\tilde{\Phi} = \int d \phi e^{i k \phi} P(\phi ) = \mathrm{tr} \left ( E^k \rho \right ) ,
\end{equation}
where the last equality holds because for all $k$
\begin{equation}
E^k   = \int d \phi e^{i k \phi} | \phi \rangle \langle \phi | ,
\end{equation}
in spite of the fact that the phase states are not orthogonal. This is to say that the lack of unitarity 
is equivalent to a description of phase in terms of a positive-operator measure. 

Despite the lack of unitarity of $E^k$ there is also a suitable Weyl form of commutation relations
\begin{equation}
\label{Wpn}
 E^k e^{i \phi \hat{n} } =   e^{i k \phi} e^{i \phi \hat{n} } E^k   .
 \end{equation}
Since the operator $E$ is not unitary, the change of $k$ by $-k$ is not  trivial, so in order to follow
the procedure in Ref. \cite{RTW} we have to construct explicitly the two Gram matrices. 

The first one for the vectors
\begin{equation}
|\Psi \rangle, \quad e^{i \phi \hat{n} } | \Psi \rangle, \quad E^{\dagger k}  |\Psi \rangle,
\end{equation}
is
\begin{equation}
G_+ = \pmatrix{1 & \Phi & \tilde{\Phi} \cr \Phi^\ast & 1 & \Omega \cr \tilde{\Phi}^\ast & \Omega^\ast & 1} ,
\end{equation}
with 
\begin{equation}
\label{detG+}
\det G_+ = 1 - \left | \Phi \right |^2 - \left | \tilde{\Phi} \right |^2 - \left | \Omega \right |^2 + \Theta + \Theta^\ast
\geq 0 ,
\end{equation}
being in this case 
\begin{equation}
\label{PhitPhi}
\Phi = \langle  \Psi | e^{i\phi \hat{n}} | \Psi \rangle, \quad \tilde{\Phi} = \langle \Psi | E^{\dagger k} | \Psi \rangle, 
\end{equation}
$\Theta = \Omega \Phi \tilde{\Phi}^\ast$, and 
\begin{equation}
\label{Om}
\Omega = \langle \Psi | e^{-i\phi \hat{n}} E^{\dagger k} | \Psi \rangle. 
\end{equation}

The second Gram matrix corresponds to the change of $\phi$ by $-\phi$ and $k$ by $-k$, so the three
vectors are now 
\begin{equation}
|\Psi \rangle, \quad e^{- i \phi \hat{n} } | \Psi \rangle, \quad E^k |\Psi \rangle,
\end{equation}
leading to
\begin{equation}
G_- = \pmatrix{1 & \Phi^\ast & \tilde{\Phi}^\ast \cr \Phi & 1 & e^{-i k \phi} \Omega^\ast \cr \tilde{\Phi} & 
e^{i k \phi} \Omega & 1 - \Pi_k} ,
\end{equation}
where Eqs. (\ref{Wpn}) and (\ref{lu}) have been used, being $\Pi_k = \langle \Psi | \hat{\Pi}_k | \Psi \rangle$ 
and the same $\Phi$, $\tilde{\Phi}$ and $\Omega$ in Eqs. (\ref{PhitPhi}) and (\ref{Om}). This leads to the 
determinant 

\begin{widetext}

\begin{equation}
\label{detG-}
\det G_- = 1 - \left | \Phi \right |^2 - \left | \tilde{\Phi} \right |^2 - \left | \Omega \right |^2 + e^{i k \phi} \Theta + 
e^{-i k \phi}  \Theta^\ast - \Pi_k \left ( 1 -   \left | \Phi \right |^2 \right ) \geq 0 ,
\end{equation}

\end{widetext}

\noindent for the same $\Theta$ above.

At this point several routes can be followed. For definiteness from now on we will focus always in the most stringent 
scenario of $k \phi = \pi$. In such a case we readily get from the sum of Eqs. (\ref{detG+}) and (\ref{detG-}) the 
following bound:
\begin{equation}
\label{com}
\left | \Phi \right |^2 + \left | \tilde{\Phi} \right |^2 + \left | \Omega \right |^2 +\frac{ \Pi_k}{2} \left ( 1 -   \left | \Phi \right |^2 
\right ) \leq 1.
\end{equation}
The lack of unitarity of the exponential of the phase reflects in the presence of the $\Pi_k$ term. Thus, whenever 
this term is absent $\Pi_k = 0$  we  recover the same expressions obtained for the 
spin-like systems. Otherwise, this term might be also moved to the right-hand side of the relation meaning 
that the nonunitarity implies a lower upper bound in accordance with the noisy nature of positive-operator
measures.

For definiteness, on what follows we will consider the following forms 
\begin{equation}
\label{Up}
U= \left | \Phi \right |^2 + \left | \tilde{\Phi} \right |^2  \leq 1, \quad 
U^\prime = \left | \Phi \right |^2 + \left | \tilde{\Phi} \right |^2 + \left | \Omega \right |^2 \leq 1, 
\end{equation} 
and
\begin{equation}
\label{Upp}
U^{\prime \prime} = \left | \Phi \right |^2 + \left | \tilde{\Phi} \right |^2 + \left | \Omega \right |^2 +
\frac{ \Pi_k}{2} \left ( 1 -   \left | \Phi \right |^2  \right ) \leq 1 ,
\end{equation}
as well as the product 
\begin{equation}
\label{V}
V = \left | \Phi \right | \left | \tilde{\Phi} \right | \leq 1/2 .
\end{equation}

\subsection{Example: phase and number states}

Readily simple examples are provided by the eigenstates of $\hat{n}$ and $E$. For the number
states $| n \rangle$ we get for all $\phi$, $k$ and $n$ that $| \Phi | = 1$,  $\tilde{\Phi}=0$, and there 
is no effect of the $\Pi_k$ term. Thus these are minimum uncertainty states. Note that we have the 
opposite conclusions regarding the uncertainty product.

On the other hand, the phase states $(\ref{ps})$ do not provide a suitable example since they are
not normalizable. Instead, we can use their normalized counterparts, that are also
eigenstates of $E$ 
\begin{equation}
\label{pc}
| \Psi \rangle = \sqrt{1- |\xi |^2} \sum_{n=0}^\infty \xi^n | n \rangle , \quad E | \xi \rangle = \xi | \xi \rangle ,
\end{equation}
with mean number of photons $\overline{n} =  |\xi |^2/(1- |\xi |^2)$. These states can be suitably 
approached in practice via quadrature  squeezed states \cite{pss}.

In this case it can be readily seen that 
\begin{equation}
\label{pc1}
\Phi = \frac{1- |\xi |^2}{1- |\xi |^2 e^{i \phi}}, \quad \tilde{\Phi} = \xi^{\ast k}, \quad \Pi_k = 1- |\xi |^{2k} ,
\end{equation}
and 
\begin{equation}
\Omega = \xi^{\ast k}  e^{-i k \phi} \frac{1- |\xi |^2}{1- |\xi |^2 e^{-i \phi}} = e^{-i k \phi} \tilde{\Phi} \Phi^\ast  . 
\end{equation}
In Fig. 1 we have represented the combinations $U$, $U^\prime$ and $U^{\prime \prime}$ in Eqs.
(\ref{Up}) and (\ref{Upp}) as functions of $| \xi |$ for $k=1$ and $\phi = \pi$. The minima of these functions 
represent maximum uncertainty and hold for phase states with very  small mean number of photons, 
i. e., $\bar{n} = 0.6, 0.7$, and  1.3, for $U$, $U^\prime$, and $U^{\prime \prime}$, respectively.  On the
other hand, when $| \xi | \rightarrow 1$, this is when $\overline{n} \rightarrow \infty$, we get  $\Phi \rightarrow 0$,  
$| \tilde{\Phi} | \rightarrow 1$, and $\Pi_k \rightarrow 0$, as expected for ideal phase states, becoming 
minimum uncertainty states.

\begin{figure}
\centering
\includegraphics[width=6cm]{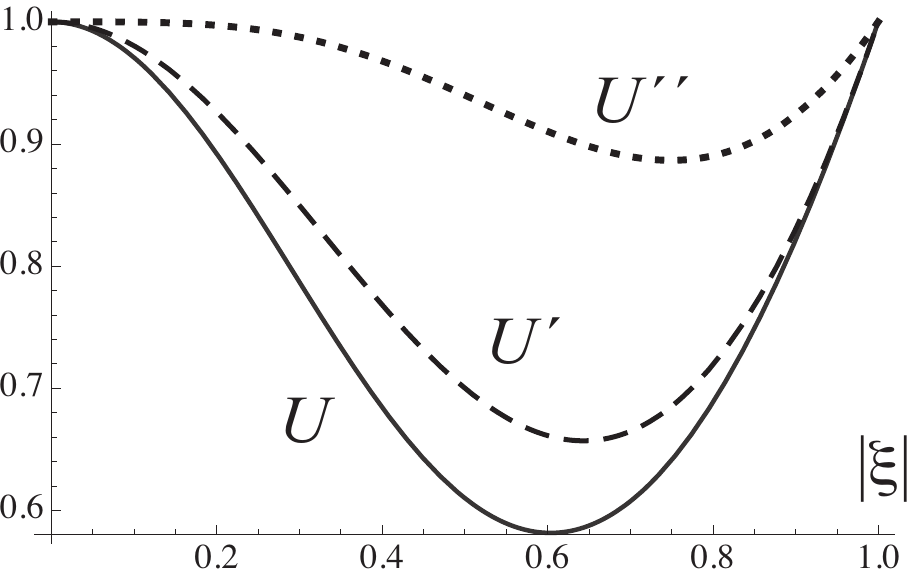}
\caption{Plot of $U$ (solid), $U^\prime$ (dashed), and $U^{\prime \prime}$ (dotted) as functions of $|\xi |$ 
for $k=1$ and $\phi = \pi$ for the phase states (\ref{pc}).}
\end{figure}

However when considering the product $V$ in Eq. (\ref{V}) again  for $k=1$ and $\phi = \pi$  it can be easily 
seen after Eq. (\ref{pc1}) that when $| \xi | \rightarrow 1$ and $| \xi | \rightarrow 0$ we get maximum uncertainty 
$V \rightarrow 0$, while $V$  attains its maximum value (i. e., minimum uncertainty),  $V=0.3$, when $| \xi | 
= 0.49$, this is $\bar{n} = 0.30$. Thus we see another clear example where maximum and minimum uncertainty 
states exchange their roles depending on the assessment of  joint uncertainty considered.

\subsection{Example: Complex Gaussian states}

States with Gaussian statistics are usually minimum uncertainty states in typical variance-based uncertainty 
relations. Then it is worth examining the case in which the number statistics can be approximated by a 
Gaussian distribution. This will work provided that the distribution is concentrated in large photon numbers 
and that it is smooth enough so that the number $n$ can be treated as a continuous variable. Thus let us 
consider a pure state $| \psi \rangle$ with 
\begin{equation}
\langle n | \Psi \rangle \simeq \left ( \frac{2a}{\pi}\right)^{1/4} \exp \left [ -(a+ib)\left ( n - \bar{n} \right )^2  \right ] ,
\end{equation}
where $\bar{n}$ represents the mean number, $a$ is given by the inverse of the number variance $\Delta^2 n = 
1/(4a)$, and $b$ provides phase-number correlations taking positive as well as negative values. Consistently 
with the above approximations we shall consider $a \ll 1$ as well as $k \ll \bar{n}$.  This situation includes the 
Glauber coherent states $| \alpha \rangle$ for large enough mean photon numbers $\Delta^2 n = \bar{n} \gg 1$ 
with $b=0$. Throughout $k\phi = \pi$ will be assumed.

In these conditions we readily get 
\begin{equation}
 \left | \Phi \right |^2  = \exp \left(-\frac{\phi^2}{4a}\right)  ,
 \quad
 \left | \tilde{\Phi} \right |^2 =  \exp\left[-\frac{(a^2+b^2)k^2}{a} \right ] ,
 \end{equation}
 and 
\begin{equation}
\left | \Omega \right |^2 =  \left | \Phi \right |^2  \left | \tilde{\Phi} \right |^2 
\exp \left ( \frac{b}{a} k \phi \right ) ,
\end{equation}
with $\Pi_k \simeq 0$.

The first thing we can notice is that $b$ increases phase uncertainty. 
Let us begin with the simplest case $b=0$. We can focus first on the plain sum relation $U$ in 
Eq. (\ref{Up}), which is plotted in Fig. 2 in solid line as a function of $a$. We can see 
that  $U $ is just a function of $ak^2$ and that minimum uncertainty, this is maximum $U$, holds for $ak^2$ 
tending both to 0 and infinity: this is when the state tends to be phase or number state, respectively, in 
accordance with the above results. In between we get a maximum uncertainty state, i. e., minimum $U$, 
when $a k^2 \simeq \pi/2$ that correspond to $ \left | \Phi \right | =  \left | \tilde{\Phi} \right |$, this is uncertainty 
equally split between phase and number. Similar results are obtained for $U^\prime$ in the same 
Eq. (\ref{Up}), as shown in Fig. 2 in dashed line.

\begin{figure}
\centering
\includegraphics[width=6cm]{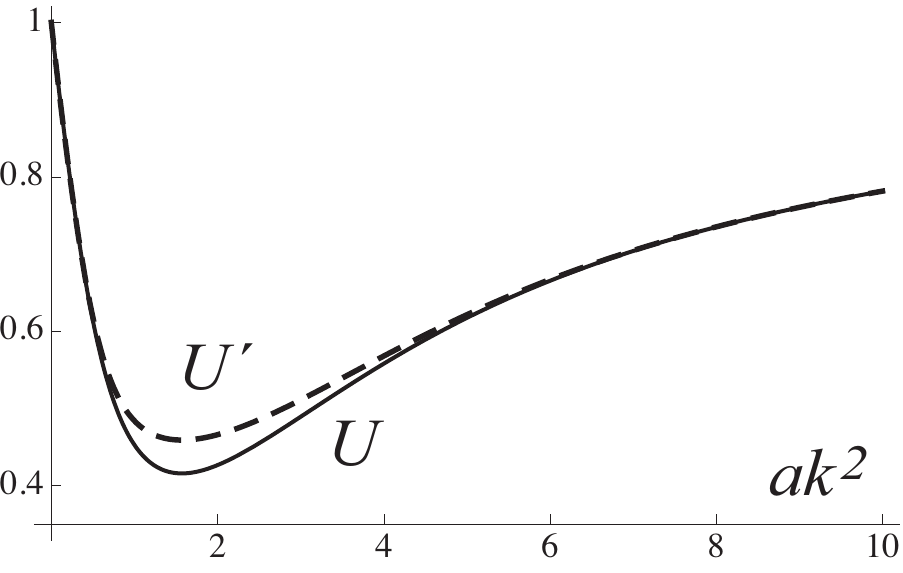}
\caption{Plot of $U$ (solid) and $U^\prime$ (dashed) as functions of $a k^2$ for Gaussian states with $b=0$.}
\end{figure}

On the other hand, the situation is quite the opposite for the certainty product $V$ in Eq. (\ref{V}):  we have 
maximum uncertainty $V \rightarrow 0$ for phase and number states $a k^2 \rightarrow 0, \infty$, while we 
have minimum  uncertainty, this is maximum $V$,  for $a k^2 = \pi/2$. This is just the opposite of the 
conclusion of the sum of characteristics.

For the case $b \neq 0$ in Fig. 3 we have plotted the certainty sums $U$ and $U^\prime$ as functions of 
$b$ for $\Delta^2 n = 10$ and $k=1$, showing that from $b=0$ increasing $b$ increases uncertainty until 
reaching $b k^2 = \pi/2$ where a revival of $U^\prime$ is produced, reaching the same certainty values 
around $b=0$. Regarding the product $V$ we have the same behavior of the case  $b=0$ but  the minimum 
uncertainty state holds for $a k^2 = (\pi/2) \sqrt{a^2/(a^2+b^2 )}$.

\begin{figure}
\centering
\includegraphics[width=6cm]{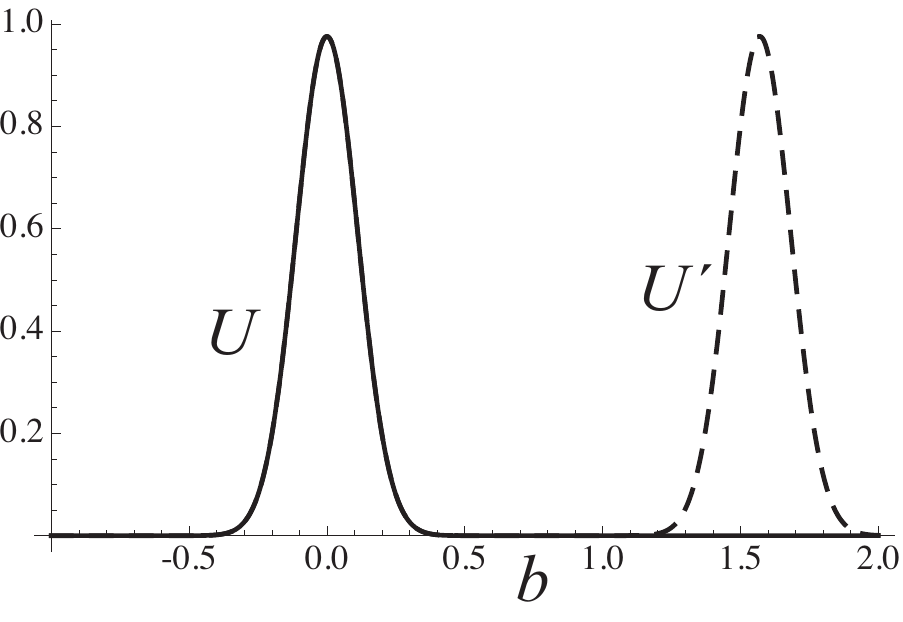}
\caption{Plot of $U$ (solid) and $U^\prime$ (dashed) as functions of $b$ for $k=1$ and Gaussian states with 
$\Delta^2 n = 10$.The two lines overlap around $b=0$.}
\end{figure}

\subsection{Example: $\hat{n}+i\lambda E^\dagger$ eigenstates}

Looking for states with interesting phase-number relations we may consider the eigenstates of $\hat{n} + 
i \lambda E^\dagger$, where $\lambda$ is a real parameter \cite{LP}:
\begin{equation}\label{eigeq}
\left ( \hat{n} + i \lambda E^\dagger \right ) | \Psi \rangle=\mu | \Psi \rangle 
\end{equation}
that has the following solution, for $\mu=0$ for definiteness,
\begin{equation}
\label{i1}
| \Psi \rangle= \frac{1}{\sqrt{I_0(2\lambda)}} \sum_{n=0}^{\infty} \frac{(-i \lambda )^n}{n!}|n\rangle ,
\end{equation}
where $I_j$ are the corresponding modified Bessel functions.

It could be interesting to apply the previous approach to these states (\ref{i1}) looking for the $\lambda$ 
that lead to minimum uncertainty. Easily we obtain:
\begin{equation}
\left | \Phi \right |^2=\frac{\left | I_0 \left( 2  \lambda e^{i \phi/2} \right) \right |^2}{I_0^2 \left(2 \lambda \right)} , 
\quad
\left | \tilde{\Phi} \right |^2 = \frac{I_k^2 \left( 2 \lambda  \right )}{I_0^2 \left( 2\lambda  \right)} ,
\end{equation}
and
\begin{equation}
\left | \Omega \right |^2 = \frac{\left | I_k \left( 2 \lambda e^{i \phi/2} \right) \right |^2 }{I_0^2 \left(2 \lambda  \right)} .
\end{equation}
As before, we focus on the case $k=1$ and $\phi=\pi$. Performing the numerical computation, we obtain 
plots of $U$ and $U'$ in Eq. (\ref{Up}) similar to the ones obtained in previous cases, as we can see in Fig. 4. 
The maximum uncertainty states are given by the minimum value of $U$ and $U'$ for a given $k$. In the 
present case the values of $\lambda$ which minimizes these functions are $\lambda=0.77$ with $\bar{n} 
= 0.8$, and $\lambda=0.88$ with $\bar{n} = 1.1$, respectively. Here again we obtain opposite results for 
the certainty product $V$ in Eq. (\ref{V}).

\begin{figure}
\centering
\includegraphics[width=6cm]{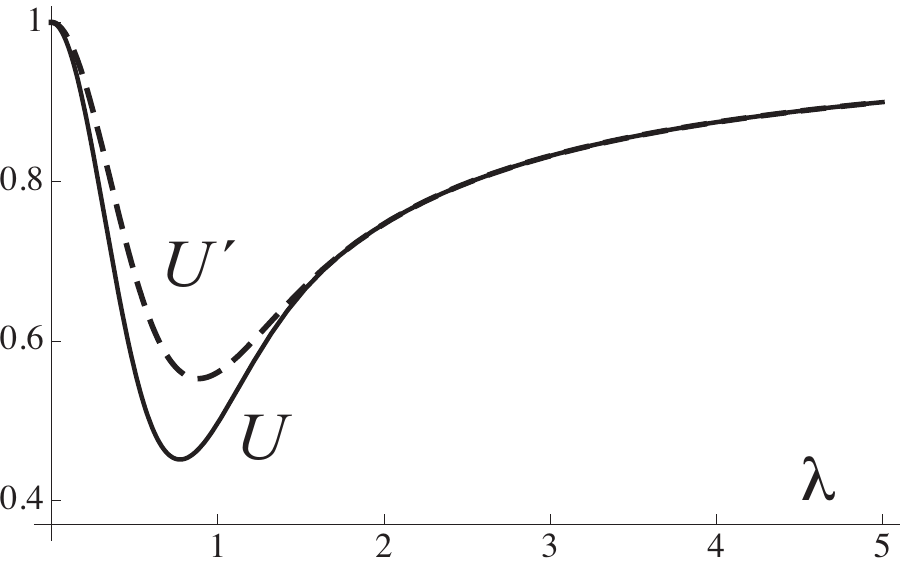}
\caption{Plot of $U$ (solid) and $U^\prime$ (dashed) as functions of $\lambda$ for the states (\ref{i1}).}
\end{figure}

\subsection{Example: Phase-number intermediate states} 

Joint uncertainty relations of two observables are often minimized by states with properties somewhat 
intermediate between the two observables. The simplest case is a readily coherent superposition of 
number and phase states of the form 
\begin{equation}
| \Psi \rangle \propto \alpha | n \rangle + \beta | \xi \rangle.
\end{equation}
We focus on the case $n >0$ and $|\xi | \rightarrow 1$ with the idea that $| \xi \rangle$ approaches the ideal 
phase states (\ref{ps}). In such a case it can be seen that the normalization condition is just $|\alpha |^2 + 
|\beta |^2 = 1$ and that
\begin{equation}
\Phi \simeq |\alpha |^2 e^{ i\phi n}, \quad 
\tilde{\Phi} \simeq |\beta |^2 \xi^{\ast k}, \quad \Omega \simeq 0 ,\quad \Pi_1 \simeq 0 ,
\end{equation}
and we shall consider $k=1$ and $\phi = \pi$.
Thus,  Eq.~(\ref{com})  reads
\begin{equation}
|\alpha |^4 +|\beta |^4 \leq 1
\end{equation}
Minimum uncertainty holds just in the limiting cases $\alpha \to 0$ and $\beta \to 0$, recovering the cases of phase 
and number states. On the other hand maximum uncertainty holds for the intermediate state $|\alpha |^2 = |\beta |^2 
= 1/2$. Clearly, the situation is reversed if we consider the product  $V$ so that maximum and minimum uncertainty 
are exchanged. 

\section{Discussion and outlook}

We have successfully derived meaningful phase-number uncertainty relations from the Weyl form 
of commutation relations. This can be applied to study phase-number statistical properties of 
meaningful field states, especially intermediate states that 
have already demonstrated interesting properties regarding uncertainty relations \cite{LP,PL,LBP}.  

Moreover, this can be a suitable tool to explore quantum metrology limits.  In typical interferometry 
$\hat{n}$ is the  generator of phase shifts. Thus the characteristic function $\Phi$ is actually expressing 
the distinguishability  of the probe state before and after a phase shift, which should be naturally related to detection 
resolution. Therefore this uncertainty relations may be connected to optimized signal detection 
schemes \cite{HVDR}.

\section*{ACKNOWLEDGMENTS}

We are grateful to \L ukasz Rudnicki for stimulating discussions, Iv\'{a}n \'{A}lvarez Domenech for his selfless help,
and Demosthenes Ellinas for valuable comments. G. D. gratefully thanks a Collaboration Grant from the spanish 
Ministerio de Educaci\'{o}n, Cultura y Deporte. A. L. thanks support from project FIS2012-35583 of spanish Ministerio 
de Econom\'{\i}a y Competitividad and from the Comunidad Aut\'onoma de Madrid research consortium 
QUITEMAD+ S2013/ICE-2801.

\end{document}